# MAARTA:Multi-Agentic Adaptive Radiology Teaching Assistant

Akash Awasthi[1], Brandon V. Chang[1], Anh M. Vu[1], Ngan Le[2], Rishi Agrawal[3], Zhigang Deng[1], Carol Wu[3], and Hien Nguyen[1]

[1] University of Houston, Houston, TX, USA
{aawasth3,bvchung,mvu9,hvnguy35}@cougarnet.uh.edu, zhigang.deng@gmail.com
[2] University of Arkansas, Fayetteville, AR, USA
thile@uark.edu
[3] MD Anderson Cancer Center, Houston, TX, USA
{RAgrawal1,ccwu1}@mdanderson.org

**Abstract.** Radiology students often struggle to develop perceptual expertise due to limited time for expert mentorship, leading to errors in visual search patterns and diagnostic interpretation. These perceptual errors—such as missed fixations, brief dwell times, or misinterpretations—are not adequately addressed by existing AI systems, which focus on diagnostic accuracy but fail to explain how and why errors occur. To bridge this gap, we propose MAARTA (Multi-Agentic Adaptive Radiology Teaching Assistant), a multi-agent framework that analyzes gaze patterns and radiology reports to provide personalized feedback. Unlike single-agent models, MAARTA dynamically recruits agents based on error complexity, ensuring adaptive and efficient reasoning. By leveraging thought graphs to compare expert and student gaze behavior, the system identifies missed findings and assigns Perceptual Error Teacher (PET) agents to analyze discrepancies. Using Chain-of-Thought (CoT) prompting, MAARTA generates meaningful insights, helping students understand their errors and refine their diagnostic reasoning, ultimately enhancing AI-driven radiology education. An anonymous code and dataset link is provided in the supplementary material.

## 1 Introduction

Radiology education requires learners to develop both technical knowledge and perceptual expertise [10,26,27]. However, a major challenge in this domain is the limited availability of expert feedback [5]. Due to demanding clinical workloads, radiologists often lack the time for personalized mentoring, leaving students without the necessary guidance to refine their diagnostic skills [2]. However, mastering radiology is not just about knowing what abnormalities look like—it is also about knowing where and how to look [17,25]. Perceptual errors in radiology are deeply connected to eye gaze behavior [7,24]. These errors often occur due to three reasons [7]: (1) a student may fail to fixate on the abnormality at all, meaning they never searched for it—similar to the "satisfaction of search" effect, where once one abnormality is found, further searching is neglected; (2) they may fixate on the abnormal region but for too short a duration, suggesting they looked but did not process the abnormality sufficiently; or (3) they may follow a reasonable gaze pattern but still miss the diagnosis due to a lack of experience or knowledge. These subtle lapses in attention allocation can lead to diagnostic mistakes, yet no AI-driven solution currently explains why these errors occur.

Recent advancements in AI, particularly Large Language Models (LLMs) and Large Multimodal Models (LMMs), offer a unique opportunity to bridge this gap. While these models have been explored for tasks such as automated report generation [22] and clinical decision support [14,21], they remain underutilized in providing personalized feedback on perceptual errors in diagnostic interpretation. Existing AI systems evaluate whether a diagnosis is correct but do not consider how a student arrived at their decision—a crucial aspect of perceptual learning. By integrating eye-tracking data with diagnostic reports, LLMs and LMMs can move beyond outcome-based assessments and offer real-time insights into how a student's visual attention compares to that of an expert, providing a more nuanced and personalized feedback mechanism. However, due to the complexity of analyzing multimodal data, particularly eye gaze patterns and radiology reports, existing systems struggle to process such information effectively. Single-agent LLM/LMM models must attempt to handle entire datasets—gaze patterns, reports, and diagnostic insights—within a single prompt. This often leads to inefficiencies and the potential loss of critical information [3].

To address this challenge, we propose MAARTA (Multi-Agentic Adaptive Radiology Teaching Assistant), a novel multi-agent LMM framework designed to analyze perceptual errors and provide personalized feedback to student radiologists. Unlike traditional single-agent AI systems, which struggle with long-context reasoning and multimodal data interpretation [19,29,18], MAARTA leverages a distributed, adaptive approach where multiple LLM/LMM agents work in parallel and independently to analyze differences between expert and student gaze patterns. This framework dynamically adjusts the number of reasoning agents based on error complexity, ensuring an efficient, scalable, and interpretable feedback mechanism. Following are the main contributions of our work.

- **Personalized Perceptual Error Feedback:** We propose a framework that analyzes a student's eye gaze data and report to explain why they missed a particular finding, offering personalized feedback to improve diagnostic skills.
- **Adaptive Multi-Agent Framework:** We introduce MAARTA, a multi-agent system that dynamically recruits LLM/LMM agents based on error complexity to process multimodal data efficiently, enhancing reasoning capabilities while maintaining scalability.
- **Simulated Perceptual Error Dataset:** We release a novel simulated dataset focused on perceptual errors in radiology, enabling further research into the reasoning capabilities of LLMs and LMMs for understanding diagnostic mistakes and improving AI-driven feedback mechanisms.

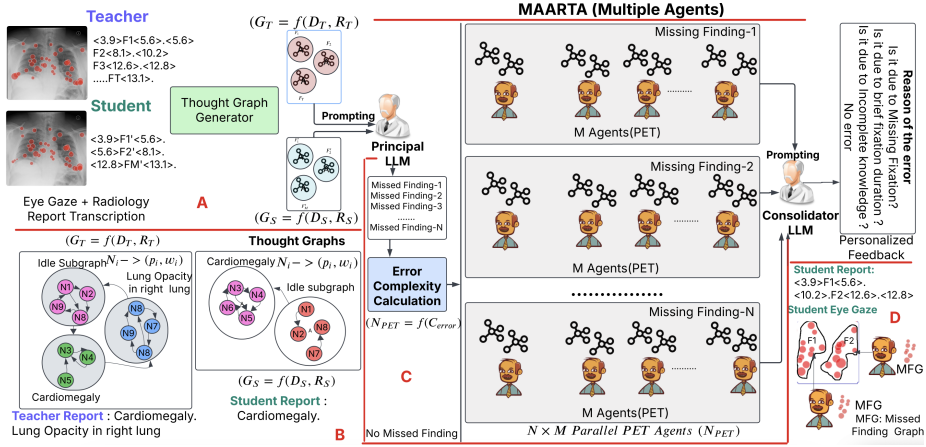

**Fig. 1.** Overview of the proposed methodology. (A) Thought graph generation process from eye gaze data. (B) Structure of the thought graphs representing expert and student gaze patterns. (C) Key modules of MAARTA for adaptive agent recruitment. (D) Workflow of PET agents in analyzing gaze sub-patterns and identifying perceptual errors.

## 2   Methodology

### 2.1   Background:

Multi-agent LLMs and LMMs represent a shift from single-model reasoning to collaborative intelligence, where multiple LLM/LMM agents work together to solve complex tasks [11,12]. Inspired by human teamwork, these systems distribute the cognitive load across specialized agents, enhancing efficiency, adaptability, and robustness [11]. Researchers have applied multi-agent LLMs and LMMs to fields such as scientific discovery [8] and medicine [16,23], where agents collaboratively address subproblems before synthesizing their insights. Multi-agent systems adopt centralized, decentralized, or hybrid architectures, balancing control and flexibility, while orchestration ensures seamless collaboration through task division, conflict resolution, and adaptability[11].

**Multi-Agent systems in Medicine:** Multi-agent LLMs and LMMs have demonstrated success in various medical applications, including radiology report generation [23], medical decision-making [16], and histopathology analysis with PathFinder [9]. These systems employ specialized agents to address specific sub-problems, integrating their insights to enhance diagnostic accuracy and decision-making

However, the use of multi-agent systems in **medical education** remains underexplored. Most studies rely on a fixed number of agents and lack methods for dynamically adjusting the agent count to optimize both accuracy and computational efficiency.

## 2.2  Problem Statement

This study explores whether an adaptive multi-agent system can provide personalized feedback to radiology students. The challenges addressed include:

- Determining the optimal number of agents for analyzing multimodal data.
- Understanding how error complexity influences the number of agents needed for effective and efficient reasoning.

Since gaze patterns and search strategies differ between learners and teachers, the number of agents should adapt based on the complexity of missed findings, optimizing computational efficiency.

## 2.3  Mathematical Formulation

We model gaze data as matrices for both the teacher (T) and student (S), represented as $D_T, D_S \in \mathbb{R}^{t \times d}$, where $t$ denotes the number of time steps and $d$ represents fixation features. The corresponding radiology reports, $R_T$ and $R_S$, are aligned with gaze fixations using a transformation function $f$, which constructs *thought graphs* $G_T$ and $G_S$. These graphs encode diagnostic reasoning as directed scene graphs, providing a more structured approach to prompt design. This process is visually represented in Figure 1A.

As illustrated in Figure 1B, each thought graph consists of nodes representing fixation points with spatial coordinates and durations, while edges define transitions between fixations. These graphs are further divided into *subgraphs*, each corresponding to a specific diagnostic finding (or thought), enabling meaningful alignment between visual attention and diagnostic interpretation.

## 2.4  Complexity-Adaptive Multi-Agent Reasoning

Inspired by [4], MAARTA adaptively determines the number of agents based on the complexity of *missed diagnostic observations*. Let $G_T = (V_T, E_T)$ and $G_S = (V_S, E_S)$ be the teacher's and student's thought graphs, with the number of subgraphs $n_T$ and $n_S$. The number of missed diagnostic findings is:

$$\Delta n = |n_T - n_S| \qquad (1)$$

Each subgraph in $G_T$ that is missed by the student is compared against all subgraphs in $G_S$, as the student's cognitive process involves all available subgraphs. Therefore, the **error complexity score** is defined as:

$$C_{\text{error}} = \Delta n \cdot n_S \qquad (2)$$

where $C_{\text{error}}$ quantifies the total number of subgraph comparisons required for reasoning. The number of agents required for reasoning is determined as a function of the error complexity:

$$N_{\text{agents}} = f(C_{\text{error}})$$

Where, $f(C_{\text{error}})$ represents a general functional relationship between error complexity and the number of agents ($N_{\text{agents}}$). For our experiments, we assume a linear relationship and recruit agents directly based on $C_{\text{error}}$. This assumption is empirically tested to evaluate its validity and explore additional influencing factors. The formulation ensures adaptive scaling of agents with task complexity, enabling parallelized reasoning and improved diagnostic accuracy. By leveraging *distributed problem-solving* principles, MAARTA balances computational efficiency and diagnostic performance.

### 2.5 MAARTA: Multi-Agentic Adaptive Radiology teaching Assiatnt

The proposed framework, MAARTA, is designed to adaptively recruit agents based on the complexity of perceptual errors in radiology education, as illustrated in Figure 1C. The framework consists of the following main components:

**Principal LLM (Global Reasoning Coordinator)**: This agent processes the student's and teacher's thought graphs $G_S = (V_S, E_S)$ and $G_T = (V_T, E_T)$, computes the difference in the number of subgraphs $\Delta n$, and identifies the missed findings. The error complexity score $C_{\text{error}}$ is then computed in the next step to determine how many agents should be recruited.

**Error Complexity Calculation (ECS)**: This step calculates the error complexity score $C_{\text{error}}$, which determines the number of PET agents. If the error complexity score $C_{\text{error}}$ is zero, then no agents are recruited, as there are no perceptual errors to analyze.

**Perceptual Error Teachers (PETs)**: For each missed finding in the student's thought graph $G_S$, a Perceptual Error Teacher (PET) agent is assigned to analyze the corresponding gaze pattern. Each PET agent focuses on a specific subgraph $g_S \subset G_S$, which represents the student's fixations and durations associated with a particular diagnostic finding. It then compares this with the teacher's corresponding gaze subgraph $g_T \subset G_T$, assessing whether the student's gaze behavior aligns with expert attention patterns.

Using Chain-of-Thought (COT) prompting, PET agents perform structured reasoning to determine if the student failed to fixate on the abnormality, exhibited brief fixation duration, or demonstrated a gaze pattern indicative of incomplete knowledge. Since different regions of an image may receive varying levels of attention based on experience, each PET agent evaluates a localized segment of the student's thought process, as shown in Figure 1D. By systematically comparing gaze distributions across subgraphs, PET agents identify perceptual discrepancies that contribute to the missed finding. Once all PET agents complete their analysis, their findings are aggregated to derive a structured explanation of the reasoning behind the student's diagnostic error. This enables an interpretable and targeted feedback mechanism, allowing the student to understand and refine their perceptual strategies.

**Consolidator LLM (Final Decision Aggregator)**: The reasoning for all missing findings is consolidated using a logical OR operation to generate the final error explanation for each case. In addition to this consolidated output, our system also provides detailed reasoning for each individual missed finding. The final result $F$ is structured as a JSON output, offering personalized feedback to the student based on the analysis of perceptual errors.

## 3   Dataset & Experimentation

**Dataset:** The **EGD-CXR** dataset [15] contains 1,083 chest X-ray (CXR) images with synchronized eye-tracking and radiology report transcription data from an experienced radiologist.

**Simulated Error Data:** Since no public dataset captures student radiologist perceptual errors, we simulate perceptual errors using EGD-CXR. Our pipeline consists of two steps: (1) *Fixation-Transcription Mapping*, where we align sentence-level timestamps from reports with gaze data, yielding 1,025 mapped samples; (2) *Error Synthesis*, introducing three error types: (i) *Missed Fixation*, where finding fixations are removed; (ii) *Reduced Fixation*, where fixation durations are halved; and (iii) *Incomplete Knowledge*, where fixations remain unchanged, but transcriptions are altered to mimic misinterpretation. This process generates a balanced dataset for evaluating MAARTA's ability to detect perceptual mistakes. Dataset statistics are detailed in the supplementary materials.

**Experiments:** We compare MAARTA against two baselines: (1) a single-agent LLM/LMM processing reports and gaze data without graph-based reasoning, and (2) a single-agent system incorporating scene graphs. Models are evaluated using zero-shot chain-of-thought (ZS-CoT) prompting across different architectures, including Llama 3.2-Instruct (3B) [6], Llama 3.2 11B-Vision-Instruct [6], Mistral 7B-Instruct-v0.3 [13], and GPT-4o [20]. To ensure consistency, we set the temperature to 0.2. GPT-4o is accessed via the OpenAI API, while Llama and Mistral models run on Together.ai [1]. We implement MAARTA using the AutoGen framework [28], enabling dynamic multi-agent coordination.

**Evaluation Metrics:** We frame this as a multilabel classification task where models must correctly classify diagnostic observations. Performance is measured using accuracy, precision, recall, F1 score, and Hamming loss. Table 1 summarizes the results, reporting mean precision, recall, and F1 scores across all classes to provide a comprehensive performance comparison.

## 4   Results & Discussion

**Quantitative Results:** Table 1 compares MAARTA with baseline models, demonstrating consistent performance gains across all metrics. Notably, GPT-4o-Mini benefits significantly from MAARTA's multi-agent reasoning, with improvements in accuracy (75.00 vs. 25.46 and 23.57) and F1 score (83.00 vs. 64.84 and 63.94) compared to its single-agent counterparts. Mistral-7B also sees improvements, though smaller, with accuracy increasing to 33.00 from 30.33 and 15.00, and F1 score rising to 53.00 from 33.59 and 38.50. Similarly, LLaMA-3.2-11B-Vision shows gains, with accuracy improving to 50.20 from 40.78 and 16.32, and F1 score increasing to 69.00 from 47.19 and 49.48. These results suggest that larger models, such as GPT-4o-Mini and LLaMA-3.2-11B-Vision, leverage distributed reasoning more effectively, while Mistral-7B benefits to a lesser extent. Despite leveraging multiple agents, MAARTA remains computationally efficient. For the GPT-4o-Mini model, the mean response time with multi-agent reasoning was 13.17 seconds, compared to 13.42 seconds for the single-agent counterpart. This slight difference highlights that MAARTA enhances reasoning capabilities without sacrificing computational efficiency. Furthermore, our results indicate that a multi-agent approach can enhance LLM/LMM capabilities in analyzing large graphs. A detailed breakdown of class-specific performance and response times is provided in the supplementary file.

Table 1. Performance comparison of different LLM/LMM models across baselines and MAARTA

| Model | Accuracy ↑ | Precision ↑ | Recall ↑ | F1 Score ↑ | Hamming Loss ↓ |
|---|---|---|---|---|---|
| **Single Agent** | | | | | |
| Mistral-7B-Instruct-v0.3-ZS CoT | 30.33 | 43.10 | 44.45 | 33.59 | 0.33 |
| GPT-4o-Mini-ZS CoT | 25.46 | 60.10 | 81.60 | 64.84 | 0.30 |
| LLAMA-3.2-11B-Vision-Instruct-ZS CoT | 40.78 | 55.17 | 50.38 | 47.19 | 0.25 |
| **Single Agent with Thought Graph** | | | | | |
| Mistral-7B-Instruct-v0.3-ZS CoT | 15.00 | 38.21 | 44.09 | 38.50 | 0.40 |
| GPT-4o-Mini-ZS CoT | 23.57 | 62.57 | 79.32 | 63.94 | 0.29 |
| LLAMA-3.2-11B-Vision-Instruct-ZS CoT | 16.32 | 50.27 | 65.94 | 49.48 | 0.39 |
| **MAARTA (Ours)** | | | | | |
| Mistral-7B-Instruct-v0.3-ZS CoT | 33.00 | 53.00 | 63.00 | 53.00 | 0.30 |
| GPT-4o-Mini-ZS CoT | 75.00 | 82.00 | 87.00 | 83.00 | 0.09 |
| LLAMA-3.2-11B-Vision-Instruct-ZS CoT | 50.20 | 62.00 | 79.00 | 69.00 | 0.19 |

**Qualitative Results:** Figure 2 illustrates a sample cases from our simulated dataset, highlighting the reasoning process of MAARTA. I-1 and I-2 are input cases, while O-1 and O-2 are the corresponding outputs. In I-1, The student fails to detect pleural effusion on both sides of the lungs, primarily due to inadequate visual attention to the affected regions. Although a single fixation point is present, it does not translate into an accurate identification of the finding. MAARTA successfully predicts the missed finding and provides an explanation, detailing the reasoning behind the omission. I-2 shows the example when student misses the finding due to not focus for enough time (duration).

**Ablation:** We conducted two ablation experiments using the GPT-4o-Mini model on our simulated error dataset: (1) adaptive PET agent assignment based on error complexity, and (2) the impact of PET agent communication.

**Adaptive PET Agents:** This experiment is performed on the complete error dataset. As shown in Table 2, dynamically assigning PET agents based on error complexity function improves performance compared to assigning a number of agents solely based on the number of errors. Figure 2 (B) illustrates the relationship between error complexity score and Hamming loss across different model sizes. While larger models exhibit a consistent trend, smaller models experience performance degradation when the number of agents increases beyond a certain threshold. This suggests that the number of agents is not a simple linear function of error complexity but also depends on model size.

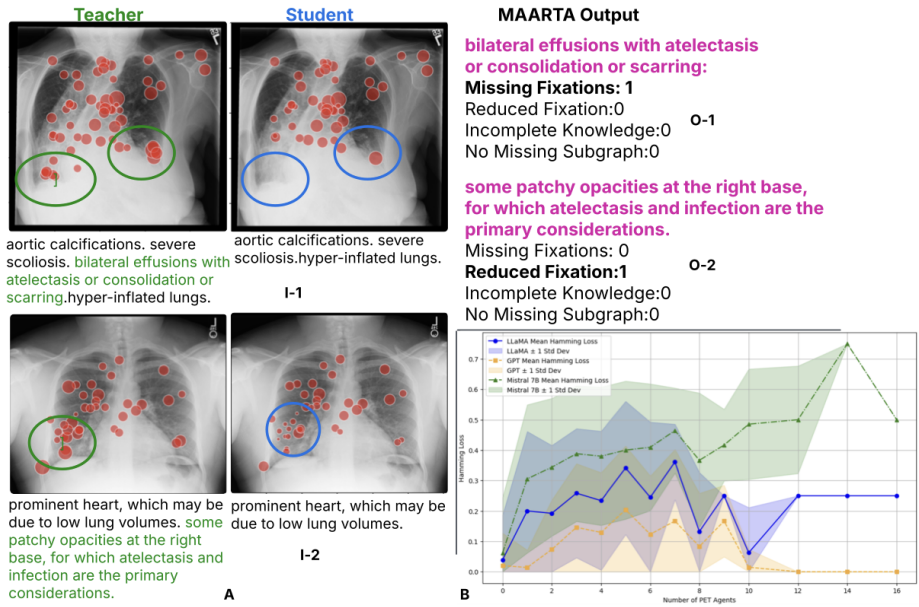

**Fig. 2.** Qualitative results: MAARTA's reasoning in detecting missed findings.

**Table 2.** Ablation study comparing different agent selection strategies and communication effects using GPT-4o.

| Adaptive Agents | | | | | |
|---|---|---|---|---|---|
| Method | Accuracy ↑ | Precision ↑ | Recall ↑ | F1 Score ↑ | Hamming Loss ↓ |
| Agents Based on Number of Errors | 66.43 | 75.00 | 78.00 | 75.00 | 0.13 |
| Agents Based on Error Complexity Function | 75.00 | 82.00 | 87.00 | 83.00 | 0.09 |
| Communication | | | | | |
| Without Communication | 66.10 | 77.00 | 89.00 | 82.00 | 0.11 |
| With Communication | 31.37 | 8.00 | 25.00 | 12.00 | 0.36 |

**PET Agent Communication:** We tested inter-agent communication by allowing PET agents to exchange information during comparison. In 100 randomly selected samples, performance dropped when agents communicated, indicating that independent PET agents are more effective for our problem.

## 5   Conclusion

MAARTA revolutionizes AI-based radiology education by offering personalized feedback on perceptual errors, shifting from passive evaluation to an active learning approach. This adaptive system enhances diagnostic training and opens the door for scalable AI-assisted medical education, enabling cognitive skill assessment and individualized mentorship at scale.

# References

1. Together ai, 2022. Accessed: 2024-11-01.
2. Robert Alexander, Stephen Waite, Michael A Bruno, Elizabeth A Krupinski, Leonard Berlin, Stephen Macknik, and Susana Martinez-Conde. Mandating limits on workload, duty, and speed in radiology. *Radiology*, 304(2):274–282, 2022.
3. K Chang, S Xu, C Wang, Y Luo, T Xiao, and J Zhu. Efficient prompting methods for large language models: A survey. arxiv 2024. *arXiv preprint arXiv:2404.01077*.
4. Lingjiao Chen, Jared Quincy Davis, Boris Hanin, Peter Bailis, Ion Stoica, Matei A Zaharia, and James Y Zou. Are more llm calls all you need? towards the scaling properties of compound ai systems. *Advances in Neural Information Processing Systems*, 37:45767–45790, 2025.
5. Cindy Chew, Patrick J O'Dwyer, and David Young. Radiology and the medical student: do increased hours of teaching translate to more radiologists? *BJR| Open*, 3(1):20210074, 2021.
6. Abhimanyu Dubey, Abhinav Jauhri, Abhinav Pandey, Abhishek Kadian, Ahmad Al-Dahle, Aiesha Letman, Akhil Mathur, Alan Schelten, and Amy Yang et al. The llama 3 herd of models, 2024.
7. Warren B Gefter, Benjamin A Post, and Hiroto Hatabu. Commonly missed findings on chest radiographs: causes and consequences. *Chest*, 163(3):650–661, 2023.
8. Alireza Ghafarollahi and Markus J Buehler. Sciagents: Automating scientific discovery through bioinspired multi-agent intelligent graph reasoning. *Advanced Materials*, page 2413523, 2024.
9. F. Ghezloo et al. Pathfinder: A multi-modal multi-agent system for medical diagnostic decision-making applied to histopathology. *arXiv preprint arXiv:2502.08916*, 2025.
10. Richard B Gunderman and Parth Patel. Perception's crucial role in radiology education. *Academic radiology*, 26(1):141–143, 2019.
11. T. Guo et al. Large language model based multi-agents: A survey of progress and challenges. *arXiv preprint arXiv:2402.01680*, 2024.
12. Shanshan Han, Qifan Zhang, Yuhang Yao, Weizhao Jin, Zhaozhuo Xu, and Chaoyang He. Llm multi-agent systems: Challenges and open problems. *arXiv preprint arXiv:2402.03578*, 2024.
13. Albert Q. Jiang, Alexandre Sablayrolles, Arthur Mensch, Chris Bamford, Devendra Singh Chaplot, Diego de las Casas, Florian Bressand, Gianna Lengyel, Guillaume Lample, Lucile Saulnier, Lélio Renard Lavaud, Marie-Anne Lachaux, Pierre Stock, Teven Le Scao, Thibaut Lavril, Thomas Wang, Timothée Lacroix, and William El Sayed. Mistral 7b, 2023.
14. Charles E Kahn Jr. Artificial intelligence in radiology: decision support systems. *Radiographics*, 14(4):849–861, 1994.
15. Alexandros Karargyris, Satyananda Kashyap, Ismini Lourentzou, Joy Wu, Arjun Sharma, Matthew Tong, Shafiq Abedin, David Beymer, Vandana Mukherjee, Elizabeth A. Krupinski, and Mehdi Moradi. Creation and validation of a chest x-ray dataset with eye-tracking and report dictation for ai development. *arXiv preprint arXiv:2009.07386*, 2020.
16. Yubin Kim, Chanwoo Park, Hyewon Jeong, Yik Siu Chan, Xuhai Xu, Daniel McDuff, Hyeonhoon Lee, Marzyeh Ghassemi, Cynthia Breazeal, and Hae Won Park. Mdagents: An adaptive collaboration of llms in medical decision making. 2024.
17. Harold L Kundel and Paul S La Follette Jr. Visual search patterns and experience with radiological images. *Radiology*, 103(3):523–528, 1972.

18. Xin Li, Weize Chen, Qizhi Chu, Haopeng Li, Zhaojun Sun, Ran Li, Chen Qian, Yiwei Wei, Chuan Shi, Zhiyuan Liu, et al. Can large language models analyze graphs like professionals? a benchmark, datasets and models. *Advances in Neural Information Processing Systems*, 37:141045–141070, 2025.
19. Xinyi Li, Sai Wang, Siqi Zeng, Yu Wu, and Yi Yang. A survey on llm-based multi-agent systems: workflow, infrastructure, and challenges. *Vicinagearth*, 1(1):9, 2024.
20. OpenAI, Josh Achiam, Steven Adler, Sandhini Agarwal, Lama Ahmad, Ilge Akkaya, Florencia Leoni Aleman, Diogo Almeida, and Janko Altenschmidt et al. Gpt-4 technical report, 2024.
21. Chintan Shah, Karapet Davtyan, Ilya Nasrallah, R Nick Bryan, and Suyash Mohan. Artificial intelligence-powered clinical decision support and simulation platform for radiology trainee education. *Journal of Digital Imaging*, 36(1):11–16, 2023.
22. Phillip Sloan, Philip Clatworthy, Edwin Simpson, and Majid Mirmehdi. Automated radiology report generation: A review of recent advances. *IEEE Reviews in Biomedical Engineering*, 2024.
23. Malavikha Sudarshan, Sophie Shih, Estella Yee, Alina Yang, John Zou, Cathy Chen, Quan Zhou, Leon Chen, Chinmay Singhal, and George Shih. Agentic llm workflows for generating patient-friendly medical reports. *arXiv preprint arXiv:2408.01112*, 2024.
24. Georgia Tourassi, Sophie Voisin, Vincent Paquit, and Elizabeth Krupinski. Investigating the link between radiologists' gaze, diagnostic decision, and image content. *Journal of the American Medical Informatics Association*, 20(6):1067–1075, 2013.
25. A Van der Gijp, CJ Ravesloot, H Jarodzka, MF Van der Schaaf, IC Van der Schaaf, Jan PJ van Schaik, and Th J Ten Cate. How visual search relates to visual diagnostic performance: a narrative systematic review of eye-tracking research in radiology. *Advances in Health Sciences Education*, 22:765–787, 2017.
26. Stephen Waite, Zerwa Farooq, Arkadij Grigorian, Christopher Sistrom, Srinivas Kolla, Anthony Mancuso, Susana Martinez-Conde, Robert G Alexander, Alan Kantor, and Stephen L Macknik. A review of perceptual expertise in radiology-how it develops, how we can test it, and why humans still matter in the era of artificial intelligence. *Academic Radiology*, 27(1):26–38, 2020.
27. Stephen Waite, Arkadij Grigorian, Robert G Alexander, Stephen L Macknik, Marisa Carrasco, David J Heeger, and Susana Martinez-Conde. Analysis of perceptual expertise in radiology–current knowledge and a new perspective. *Frontiers in human neuroscience*, 13:213, 2019.
28. Qingyun Wu, Gagan Bansal, Jieyu Zhang, Yiran Wu, Beibin Li, Erkang Zhu, Li Jiang, Xiaoyun Zhang, Shaokun Zhang, Jiale Liu, et al. Autogen: Enabling next-gen llm applications via multi-agent conversation. *arXiv preprint arXiv:2308.08155*, 2023.
29. Yusen Zhang, Ruoxi Sun, Yanfei Chen, Tomas Pfister, Rui Zhang, and Sercan Arik. Chain of agents: Large language models collaborating on long-context tasks. *Advances in Neural Information Processing Systems*, 37:132208–132237, 2025.